\documentclass[12pt,preprint]{aastex}

\usepackage{graphicx}
\usepackage{textcomp}
\usepackage{epstopdf}

\newcommand{\permil}{\textperthousand}
\newcommand{\delval}[2]{$\delta^{#1}\textrm{#2}$}
\newcommand{\iso}[2]{$^{#1}\textrm{#2}$}
\newcommand{\gcm}{$\textrm{g}/\textrm{cm}^{3}$}

\slugcomment{ApJ Letters}

\begin{document}
\title{C, N, and O isotopic heterogeneities in low-density supernova graphite grains from Orgueil}
\author{Evan Groopman, Thomas Bernatowicz, and Ernst Zinner}
\affil{Washington University}
\affil{One Brookings Drive, Campus Box 1105, Saint Louis, MO 63130}
\email{eegroopm@physics.wustl.edu}

\begin{abstract}
\par We report on the results of NanoSIMS isotope imaging of low-density supernova graphite grains from the Orgueil meteorite.  70nm-thick microtomed sections of three supernova graphite grains were deposited on Si wafers and isotopically imaged in the NanoSIMS.  These sections contain hotspots of excesses in \iso{18}{O} and \iso{15}{N}, which are spatially well-correlated, and are likely carried by internal TiC subgrains.  These hotspots are considerably more enriched in \iso{18}{O} and \iso{15}{N} than the host graphite grain.  Correlations between \iso{18}{O} and \iso{15}{N} excesses indicate that the grains incorporated material from the He/C supernova zone.  Isotope images of the surfaces of some grains show heterogeneities in their N and O isotope compositions, with extreme excesses in \iso{15}{N} and \iso{18}{O}.  In the microtome sections we also observe two types of heterogeneities in the grains' C isotopic compositions: smooth, radial gradients in \iso{12}{C}/\iso{13}{C}, with this ratio trending towards solar with increasing radius; and highly anomalous pockets up to 2\micron\ in size with \iso{12}{C}/\iso{13}{C}$>>$solar that are located near the centers of the grain sections.  Partial isotopic equilibration does not likely explain the C isotopic heterogeneities.  These grains and their constituent parts probably formed in a stellar environment with changing isotopic composition.  
\end{abstract}
\keywords{Supernovae: General --- meteorites, meteors, meteoroids}
 
\section{Introduction}
\par Low-density graphite grains ($\rho\sim$1.7 \gcm) have been isolated from the Murchison CM2 \citep{amari1994} and Orgueil CI \citep{jadhav2006} carbonaceous chondrites.  Isotopic signatures of most low-density graphite grains indicate an origin in Type II supernovae, specifically excesses in \iso{15}{N}, \iso{18}{O}, \iso{28}{Si}, and the inferred presence of \iso{26}{Al} and \iso{44}{Ti} from excesses in \iso{26}{Mg} and \iso{44}{Ca}, respectively, relative to solar composition \citep{amari1995c,travaglio1999,stadermann2005,jadhav2006}.  Low-density graphite grains also exhibit a wide range of \iso{12}{C}/\iso{13}{C} ratios.  \citet{stadermann2005} obtained nano-scale secondary ion mass spectrometry (NanoSIMS) isotope images in \iso{12,13}{C}, \iso{16,17,18}{O}, and \iso{46,47,48,49,50}{Ti} of 9 microtome sections of the graphite spherule KE3e\#10 \citep{croat2003} from the Murchison low-density KE3 fraction ($\rho$ = 1.65-1.72\gcm).  These microtomed graphite slices contained gradients in the \iso{12}{C}/\iso{13}{C} and \iso{16}{O}/\iso{18}{O} isotopic ratios, with their interiors being more anomalous than their exteriors.  It was suggested that these gradients are due to partial isotopic equilibration with the grain's environment during its lifetime.  Strong initial O$^-$ secondary ion signals with respect to the ``bulk'' graphite were found in small regions that contain to internal TiC subgrains.  The TiC subgrains are more enriched in \iso{18}{O} than the parent grain, although their C isotopic compositions are indistinguishable from those of the surrounding graphite.

\par We report on the results of NanoSIMS isotope imaging of low-density supernova graphite grains from the Orgueil meteorite.  We acquired isotope images of the surfaces of numerous grains as well as isotope images of microtomed sections of their interiors.  Isotopic imaging is a useful tool for investigating the heterogeneous isotopic composition within presolar grains.  This study is part of correlated isotopic, chemical, and crystallographic investigations undertaken to characterize individual presolar grains as fully as possible.

\section{Samples and Experimental Methods}
\par  Seven size/density fractions were isolated from the Orgueil CI chondrite (OR1b-OR1h)\footnote{OR1b label: OR = Orgueil, 1 = size $>$1\micron, b = density ($\rho$ = 1.59-1.67 \gcm)} with densities increasing alphabetically \citep{jadhav2006}.  The OR1d fraction ($\rho$ = 1.75-1.92 \gcm) contains considerable insoluble organic matter of solar isotopic composition, which had not been removed during the chemical separation, and which surrounds the graphite grains.  A few $\sim$0.5$\mu$L drops of OR1d suspended in a 4:1 mixture of isopropanol/water were pipetted onto high-purity Au foil and briefly dessicated in a vacuum oven at 70\textcelsius\ \citep{eeg2011}.  These graphite grains were identified in the scanning electron microscope (SEM) based upon surface morphology and Energy Dispersive X-Ray Spectroscopy (EDXS).  Candidate grains were picked with a sharp W micromanipulator needle and were deposited on a clean Au-foil mount to reduce contamination from the insoluble organic matter; some of this material, however, remains stuck to the grains' surfaces.  Twenty candidate grains with diameters $>$5\micron\ were selected, labeled OR1d6m\footnote{6m = the 6$^{th}$ mount of OR1d}.  Raman microprobe spectra with 532nm excitation and $\sim$1\micron\ resolution were obtained for all grains \citep{wopenka2011}.

\par A several-nanometer-thick coating of Pt was deposited in a sputter coater on the grains to secure them to the Au foil.  Whole-grain NanoSIMS measurements of negative secondary ions (\iso{12,13}{C}, \iso{14,15}{N}, \iso{16,18}{O}, and \iso{28,29,30}{Si}) were made with a Cs$^+$ primary beam ($\sim$100nm), and measurements of positive secondary ions (\iso{24,25,26}{Mg}, \iso{27}{Al}, \iso{28,29,30}{Si}, \iso{39,41}{K}, \iso{40,42,43,44}{Ca}, \iso{46,47,48,49,50}{Ti}, \iso{51}{V}, and \iso{52}{Cr}) were made with an O$^-$ primary beam.  To obtain the N isotopic composition we measured \iso{12}{C}\iso{14}{N}$^-$ and \iso{12}{C}\iso{15}{N}$^-$, as N does not have a stable negative ion.  We used sufficiently high mass resolution to separate the isobaric peaks of \iso{12}{C}\iso{15}{N} and \iso{13}{C}\iso{14}{N}; and \iso{12}{C}\iso{14}{N} and \iso{13}{C}\iso{13}{C} and \iso{12}{C}\iso{13}{C} H.  Each grain's bulk isotopic composition was measured by rastering the primary beam over the entire grain's surface and summing the counts for each isotopic species.  Additionally, we acquired 256 pixel by 256 pixel isotope images of the surfaces of grains G6, G13, G17, G21 in \iso{12}{C}, \iso{14,15}{N}, \iso{16,18}{O} and of G24 in \iso{12,13}{C}, \iso{28,29,30}{Si}.

\par After NanoSIMS analysis, three grains (G17, G18, G24) with sizes 18, 12, and 13\micron\ were picked from the Au foil mount, embedded in LR White hard acrylic resin, and ultramicrotomed into 70nm-thick sections. Grains were selected based upon their isotopic compositions and relatively large sizes, (Table \ref{or1d6m_data}).  Slices of each grain were deposited on Si wafers and on Cu transmission electron microcope (TEM) grids, both SiO and holey-C coated.  Due to the grains' large sizes, we were able to obtain hundreds of microtome slices of each grain and distribute these among the 3 different substrates.  Each substrate has its own advantage: holey-C substrates are useful for determining Si content and performing EELS over holes; the SiO substrate reduces the C background necessary for C K-edge X-ray absorption near-edge structure (XANES) measurements \citep{nittler2011}; and the Si wafers, while precluding TEM studies, are a superior heat sink for Raman spectroscopy (high signal-to-noise ratio), and provide a highly stable substrate for NanoSIMS measurements.  Due to the large sizes of the grains, many of the slices contain tearing holes from microtoming, shown as grey in the figures.  For the slices on Si wafers, isotope images in \iso{12}{C}, \iso{14,15}{N}, \iso{16,18}{O} or \iso{12,13}{C}, \iso{28,29,30}{Si} were obtained with the resin and Si wafer used as isotopic standards.  Only one image was obtained for each slice.  Each isotope image consists of 20 layers of a 256 pixel by 256 pixel raster area, with a dwell time per pixel per layer of 10ms.  Each raster area is chosen to be 1-2\micron\ wider than the grain slice.  While isotope images require much longer measurement times than bulk analyses to gather meaningful statistics, they allow us to investigate heterogeneities in the grains' isotopic compositions.

\section{Results}
\subsection{Whole-grain Isotopic Measurements}
\par  Whole-grain isotopic measurements are presented in Table \ref{or1d6m_data} \citep{eeg2011}.  The majority of grains have signatures of a supernova origin. Six of the 20 grains from OR1d6m are not included, as their isotopic and Raman spectroscopic measurements do not identify them as presolar grains. Each grain's measured bulk composition represents the lower bound of the grain's deviation from solar composition due to surface contamination from insoluble organic material.  From isotope images of the grains' surfaces we can locate regions of low contamination and infer a more correct isotopic composition.  We must compromise between obtaining accurate isotopic ratios via long measurements during which surface contamination is removed by sputtering and saving as much of the grain as possible for future correlated TEM, XANES, and Raman studies.

\placetable{or1d6m_data}

\subsection{Surface Images}
\par
\begin{figure}[ht]
\centering
  \epsscale{.6}
  \plotone{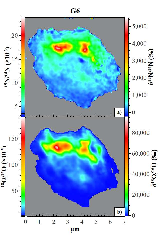}
  \figcaption{False-color isotope images of the surface of G6. Images are scaled in absolute ratios (left) and $\delta$-values (right) where $\delta$\iso{i}{X} =  $\delta$\iso{i}{X}/\iso{j}{X} = 1000$\times\big($(\iso{i}{X}/\iso{j}{X})$_{sample}$/(\iso{i}{X}/\iso{j}{X})$_{standard}$ - 1$\big)$ and \iso{j}{X} is the major isotope \iso{14}{N} or \iso{16}{O}. Black bars on left indicate terrestrial isotopic ratios (\iso{15}{N}/\iso{14}{N}$_{terrestrial}$ $\approx$ 0.004; \iso{18}{O}/\iso{16}{O}$_{terrestrial}$ $\approx$ 0.002). a) \delval{15}{N} isotope image. Deviations from terrestrial range up to 6,400 \permil\ (\iso{15}{N}/\iso{14}{N}=0.03, \iso{14}{N}/\iso{15}{N}=37).  b) \delval{18}{O} isotope image. Deviations from terrestrial range up to 98,000 \permil\ (\iso{18}{O}/\iso{16}{O}=0.20, \iso{16}{O}/\iso{18}{O}=5). There is a good spatial correlation between the hotspots in \delval{18}{O} and \delval{15}{N}.\label{G6}}
\end{figure}

 Isotopic imaging on the graphite surfaces show heterogeneities in their C, O, and N isotopic ratios.  While regions of roughly solar C, O, and N isotopic composition may be explained due to surface contamination, isotopically anomalous regions are found on the grains' surfaces, with extreme excesses in \iso{18}{O} and \iso{15}{N} relative to solar ratios.  On G6, G13, and G21 these anomalous regions in have a good spatial correlation.  If expressed as delta values (see Fig. \ref{G6} for definition), \delval{15}{N} and \delval{18}{O}, respectively, range up to 6,400\permil\ and 98,000\permil\ (\iso{15}{N}/\iso{14}{N}=0.03 and \iso{18}{O}/\iso{16}{O}=0.20; \iso{14}{N}/\iso{15}{N}=37 and \iso{16}{O}/\iso{18}{O}=5) in G6; 2,300\permil\ and 49,000\permil\ (0.012 and 0.1; 82 and 10) in G13; and 2,300\permil\ and 98,000\permil\ (0.012 and 0.20; 82 and 5) in G21.  In G17 the anomalous regions on the surface in \delval{15}{N} and \delval{18}{O} range up to 15,000\permil\ and 15,000\permil\ (.06 and .03; 17 and 31), respectively, but are not spatially correlated.  Hotspots on the grains' surfaces vary in size from 110nm to 425nm. 

\placefigure{G6}

\subsection{Microtome Sections}
\par
\begin{figure}[ht]
  \centering
  \epsscale{.6}
  \plotone{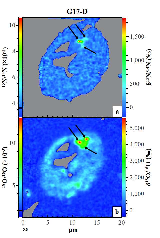}
  \figcaption{False color images of grain G17 slice D in \delval{15}{N} (a) and
\delval{18}{O} (b). There is a strong spatial
correlation between  \delval{15}{N} and
\delval{18}{O} in three hotspots, indicated by arrows. Grey
regions indicate
holes in the slice from microtoming.\label{G17all}}
\end{figure}

\begin{figure}[ht]
  \centering
  \epsscale{1}
  \plotone{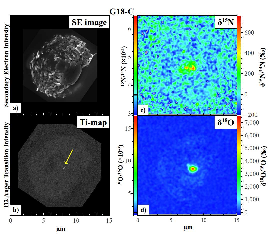}
  \figcaption{a) Auger Nanoprobe secondary electron image of G18 slice C
after NanoSIMS imaging. b) Auger Ti elemental map. A Ti-rich subgrain is clearly visible in the center of the slice, indicated by arrow. c) False-color N isotope ratio image. \delval{15}{N} up to 760\permil\ (\iso{15}{N}/\iso{14}{N}=0.007, \iso{14}{N}/\iso{15}{N}=154). d) False-color O isotope ratio image. \delval{18}{O} up to 7300\permil\ (\iso{18}{O}/\iso{16}{O}=0.017, \iso{16}{O}/\iso{18}{O}=60). There is a good spatial correlation between hotspots in panels (b)-(d). All images have the same scale and same orientation.\label{G18C}}
\end{figure}

 Multiple ultramicrotome sections of G17 and G18 contain highly anomalous and spatially-correlated hotspots of \delval{18}{O} and \delval{15}{N} [Figs. \ref{G17all},\ref{G18C}].  Five hotspots were found in 5 slices of G17 and G18 imaged in O and N isotopes.  \delval{15}{N} and \delval{18}{O} in the hotspots range up to 4400\permil\ and 6100\permil\ (\iso{15}{N}/\iso{14}{N}=0.02 and \iso{18}{O}/\iso{16}{O}=0.01; \iso{14}{N}/\iso{15}{N}=50 and \iso{16}{O}/\iso{18}{O}=70), respectively.  There is a strong spatial correlation between the observed excesses in \iso{18}{O} and \iso{15}{N}. It is likely that internal subgrains, particularly TiC, are the carriers of the isotopic anomalies in O and N within these hotspots.  Preliminary TEM studies have confirmed the presence of internal TiC subgrains in other slices of G17, G18, and G24, although no \delval{18}{O}, \delval{15}{N} hotspots were found in the two slices of G24 imaged in O and N isotopes.  Auger Nanoprobe scans of slice G18-C after NanoSIMS imaging show a strong Ti signal at the location of the \delval{15}{N}, \delval{18}{O} hotspots [Fig. \ref{G18C}b].  The four other slices with \delval{18}{O} and \delval{15}{N} hotspots were sputtered away completely during NanoSIMS imaging; we found no elemental Ti traces in Auger scans of these sections.  We did not observe differences between the \iso{16}{O}$^-$ and \iso{12}{C}\iso{14}{N}$^-$ signals between any of these hotspots and the surrounding material.  Therefore the anomalous $\delta$-value compositions of these hotspots are characterized by excesses in \iso{18}{O} and \iso{15}{N} and not by depletions in the major isotopes, \iso{16}{O} and \iso{14}{N}, respectively.  Si isotope images show uniform isotopic distributions within the grains and yield no evidence of internal SiC grains within these slices.

\placefigure{G17all}
\placefigure{G18C}

\notetoeditor{I would like figures 3 and 4 as large as possible since it is multi-paneled and in color.}
\notetoeditor{I also have these 4-panel figures as columns (4x1) instead of (2x2) if that would work better.}

\par
\begin{figure}[ht]
  \centering
  \epsscale{.9}
  \plotone{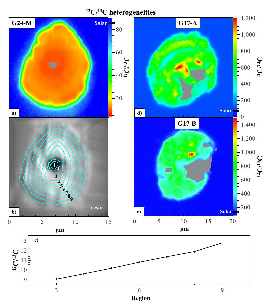}
  \figcaption{a) G18 slice M exhibits a radial gradient in
\iso{12}{C}/\iso{13}{C} from 8 to 14. b) Contours used
in (c). c) Statistically significant trend in mean \iso{12}{C}/\iso{13}{C} within concentric regions outlined in (b). All points have 5 $\sigma$ error bars in \iso{12}{C}/\iso{13}{C} based on counts. This grain likely formed in an environment where the C isotopic composition was changing on the timescale of grain formation. d) \& e) G17 slices A and B, respectively. These slices have heterogeneously distributed C isotopic compositions with highly anomalous pockets with sizes up to 2\micron\ where \iso{12}{C}/\iso{13}{C}$>>$solar. \iso{12}{C}/\iso{13}{C} varies from 400 to 1800. These pockets are most likely presolar graphites that formed earlier and were incorporated into the host grain.\label{C-hetero}}
\end{figure}

 The interiors of the graphite grains contain C isotopic heterogeneities of two forms.  Slices M and N of G24 exhibit fairly smooth gradients with \iso{12}{C}/\iso{13}{C} increasing radially from 8 to 14 [Fig. \ref{C-hetero}a-c]. Based upon the \iso{12}{C}/\iso{13}{C} ratio measured in concentric shells of the slices, there exists a statistically significant gradient in the grain's C isotopic composition. In slices G17-A and G17-B [Fig. \ref{C-hetero}d-e] \iso{12}{C}/\iso{13}{C} ranges from 400 to 1800, whereas the bulk surface isotopic measurement was 340.  The largest anomalies lie in pockets up to $\sim$2\micron\ in size close to the center of the slice.  These pockets have the same Si isotopic composition and \iso{28}{Si}$^-$
count rate as the rest of the graphite grain slice, again yielding no evidence for the presence of internal SiC subgrains.

\placefigure{C-hetero}

\section{Discussion and Conclusions}
\par We have found evidence for heterogeneities in the C, N, and O isotopic compositions of 6 low-density graphite grains.  Several of these grains contain bulk excesses of \iso{18}{O} and \iso{15}{N}, evidence of a supernova origin; however, they also contain spatially-correlated isotopic anomalies on their interiors and exteriors that are much more extreme than the bulk graphite measurements.  TiC subgrains are a likely carrier of the observed \delval{18}{O} and \delval{15}{N} heterogeneities in the grains' interiors.  Previous NanoSIMS studies of microtomed slices of a supernova graphite found that internal TiC subgrains have much larger \iso{18}{O} excesses than the parent grain \citep{stadermann2005}.  Initial \iso{16}{O}$^-$ signals associated with these internal TiCs were also much higher than those in the parent grain.  It is unknown whether the higher initial O$^-$ secondary signal is due to intrinsically higher O content in the TiC subgrains or whether O ion yields from TiC are greater than those from graphite.  However, recent electron energy loss spectroscopy (EELS) of supernova TiC subgrains has shown that they contain amorphous rims \citep{daulton2009} enriched in O up to 20 at.\% relative to the interiors of the TiC subgrains \citep{daulton2012}.  These enrichments thus may be due to O-rich amorphous rims around the TiC grains. In the present study, we do not observe higher initial O$^-$ secondary signals from the \delval{18}{O} and \delval{15}{N} hotspots.  During our NanoSIMS measurements we presputtered the graphite sections until we achieved a constant \iso{12}{C}$^-$ signal before we began imaging; this may have sputtered away the top surface of any O-rich rims on TiCs in contrast to \citet{stadermann2005}, who did not presputter their slices before imaging.

\par These spatially-correlated \delval{18}{O} and \delval{15}{N} hotspots suggest that the internal TiC grains in the supernova outflow formed from inner He/C zone material where \iso{18}{O}/\iso{16}{O} and \iso{15}{N}/\iso{14}{N} are very high \citep{rauscher2002}, while the bulk of the host graphite grain formed from a mixture that included material further out in the He/C zone or from other zones, resulting in more modest \iso{18}{O} and \iso{15}{N} excesses. The high \iso{12}{C}/\iso{13}{C} ratios of G17 and G18 also indicate contributions from the He/C zone, which has \iso{12}{C}/\iso{13}{C}$>>$solar.  While it has been proposed that carbonaceous grains form in a supernova environment where the elemental ratio C/O$<$1 \citep{clayton2011}, all of our observations point to major He/C zone contributions, where C/O$>$1.  G24 has a very low \iso{12}{C}/\iso{13}{C} ratio and relatively smaller excesses of \iso{18}{O} and \iso{15}{N}, indicating that it may represent a mixture of material with a significant contribution from the He/N zone, where C/O is still $>$1.  We did not observe any spatially-correlated hotspots in \delval{18}{O} and \delval{15}{N} in G24, however TEM investigations have found a large 300nm TiC grain in another slice on a TEM grid.  Since this slice is part of an ongoing TEM study, we will measure the O and N isotopic composition of this TiC in the future. 

\par Isotope images of grain surfaces yield some of the largest excesses in
\iso{18}{O} and \iso{15}{N} ever measured in presolar graphite grains.  Although TiC abundance has been shown to decrease with radial position in several supernova graphites \citep{croat2003}, TiC subgrains are not precluded from being the source of the surface \delval{18}{O}, \delval{15}{N} hotspots.

\par We observed two types of heterogeneities in the C isotopic compositions of microtomed graphites: smooth radial gradients in G18 and G24, and extremely anomalous pockets near the center of grain G17.  Both heterogeneities are statistically significant.  Radial gradients in C and O isotope composition have been previously observed \citep{stadermann2005}.  While partial isotopic equilibration with grains' environments may explain the O isotope compositions, gradients in C isotope composition require a different explanation.  Since the graphite grains are overwhelmingly composed of carbon, partial isotopic equilibration would require the exchange of a significant portion of the grain's mass over its lifetime in order to reproduce the gradients we observe, which is unlikely.  The equilibration of C isotopes would simultaneous have to preserve N and O isotopic anomalies that we observe.  Additionally, \citet{zinner2012} found that the radiogenic components of \iso{26}{Mg}, and \iso{41}{K} within graphite grains from the same Orgueil size/density fraction are spatially correlated with the major isotopes of the parent species in each system, implying that radiogenic daughter isotopes are quantitatively retained and there is no redistribution within the grains.  We therefore find it unlikely that significant redistribution or exchange of C has occurred.  Instead we suggest that the grains condensed in an environment whose isotopic composition was changing during grain formation.  This would likely require small-scale mixing of material in the grains' formation environments on formation timescales and/or movement of the growing grain through regions of changing isotopic composition.  Evidence of extinct \iso{49}{V} ($t_{1/2}$ = 330 days) in SiC-X grains implies that the timescale of SiC formation in supernova ejecta is on the order of several months \citep{hoppe2002}.  However, excluding very high pressure environments, graphite is likely to form before SiC from a gas with C/O $>1$ and otherwise approximately solar composition \citep{bernatowicz1996}.  Such conditions indicate that formation of the graphite grains occurred within the first few months after the supernova explosion.

\par The heterogeneous \iso{12}{C}/\iso{13}{C} anomalies in G17 [Fig. \ref{G17all}d,e] range up to \iso{12}{C}/\iso{13}{C} = 1800 and are located in distinct regions within the slices.  These anomalous regions are unlikely to be due to the presence of  either internal TiC or SiC X grains; their sizes range up to 2\micron, whereas the largest internal TiC found to date was $\sim$500nm; they lack significant Si and Ti elemental signals with respect to the rest of the graphite slice as measured in the Auger Nanoprobe; and they show no significant enrichments in \iso{28}{Si} with respect to the rest of the graphite in NanoSIMS images.  We therefore believe that the pockets are graphite regions that formed from more \iso{12}{C}-enriched material before being incorporated by the parent graphite.

\acknowledgments{This work was funded, in part, by NASA Earth
and Space Sciences Fellowship (NESSF) NNX11AN60H , and NASA grants NNX10AI45G, and NNX11AH14G. 
We thank Frank Gyngard and Kevin Croat for their helpful discussions.}

\clearpage
\begin{deluxetable}{lcccccccc}
\tabletypesize{\scriptsize}
\rotate
\tablecaption{OR1d6m Presolar Graphite: Whole-grain isotopic measurements. \label{or1d6m_data}}
\tablewidth{0pt}
\tablehead{
\colhead{Grain} &\colhead{Size ($\micron$)} &
\colhead{\iso{12}{C}/\iso{13}{C}} &
\colhead{\iso{16}{O}/\iso{18}{O}} &
\colhead{\iso{14}{N}/\iso{15}{N}} &
\colhead{\delval{29}{Si} (\permil)\tablenotemark{\dag}} &
\colhead{\delval{30}{Si} (\permil)} &
\colhead{\big(\iso{26}{Al}/\iso{27}{Al}\big)$\times$10$^{3}$} &
\colhead{\big(\iso{44}{Ti}/\iso{48}{Ti}\big)$\times$10$^{3}$}
}
\startdata
G-4	&6	&{\bf7.5} $\pm$ 0.1	&486.0 $\pm$ 9.2	&{\bf236.0} $\pm$ 18.3	&{\bf -39} $\pm$ 11	&{\bf-62} $\pm$ 13	&\nodata	&{\bf106} $\pm$18.7 \\
G-5	&10.5	&{\bf45.9} $\pm$ 0.3	&{\bf323.5} $\pm$ 5.7	&{\bf248.0} $\pm$ 10.7	&{\bf -74} $\pm$ 12	&{\bf-135} $\pm$ 15	&{\bf17.7} $\pm$ 0.6	&{\bf5.8} $\pm$ 1.0 \\
G-8	&8	&{\bf127.8} $\pm$ 0.9	&{\bf280.4} $\pm$ 5.0	&{\bf145.0} $\pm$ 9.3	&{\bf48} $\pm$ 10	&{\bf -29} $\pm$ 13	&{\bf77.1} $\pm$ 0.8	&{\bf27.3} $\pm$ 1.7 \\
G-9	&5.5	&{\bf8.0} $\pm$ 0.1	&{\bf338.5} $\pm$ 8.5	&{\bf233.8} $\pm$ 12.0	&-37 $\pm$ 40	&{\bf159} $\pm$ 55	&{\bf18.4} $\pm$ 0.7 	&{\bf109} $\pm$ 49 \\
G-11	&7.5	&{\bf22.0} $\pm$ 0.2	&{\bf297.7} $\pm$ 6.0	&{\bf232.2} $\pm$ 10.3	&-21 $\pm$ 11	&{\bf-40} $\pm$ 14	&{\bf73.6} $\pm$ 0.9 	&{\bf3.4} $\pm$ 0.8 \\
G-15	&12	&{\bf13.4} $\pm$ 0.1	&{\bf471.8} $\pm$ 10.5	&{\bf240.2} $\pm$ 16.4	& -21 $\pm$ 24	& 7 $\pm$ 30	& \nodata  	& \nodata \\
G-20	&4.5	&{\bf74.3} $\pm$ 0.5	&501.6 $\pm$ 10.3	& 273.9 $\pm$ 16.9	&{\bf-37} $\pm$ 14	&-32$\pm$ 18	&\nodata  	&{\bf46.2} $\pm$ 6.1 \\
G-23	&6	&{\bf73.7} $\pm$ 0.5	&{\bf394.4} $\pm$ 7.2	&{\bf214.4} $\pm$ 13.1	&{\bf-134} $\pm$ 8	&{\bf-168} $\pm$ 10	&{\bf68.9} $\pm$ 0.7 &{\bf19.5} $\pm$ 3.5 \\	
\tableline								
\sidehead{Imaged Grains:}								
G-6	&4.5	&{\bf113.2} $\pm$ 0.8	&{\bf163.2} $\pm$ 2.8	&{\bf145.3} $\pm$ 5.4	&-13 $\pm$ 13	&{\bf-60} $\pm$ 16	&{\bf446} $\pm$ 1.5	&{\bf103} $\pm$ 6.0 \\
G-13	&5.5	&{\bf66.0} $\pm$ 0.4	&{\bf170.1} $\pm$ 3.2	&263.0 $\pm$ 5.8	&{\bf-141} $\pm$ 17	&{\bf-181} $\pm$ 21	&{\bf231} $\pm$2.4 	&{\bf349} $\pm$ 55.9 \\
G-17	&18	&{\bf306.1} $\pm$ 2.1	&{\bf306.3} $\pm$ 6.4	&{\bf215.1} $\pm$ 10.4	&-27 $\pm$ 18	&{\bf-86} $\pm$ 21	&{\bf55.8} $\pm$ 2.0 	&{\bf130} $\pm$ 13.5 \\
G-18	&12	&{\bf121.0} $\pm$ 0.8	&{\bf432.9} $\pm$ 9.2	&{\bf148.8} $\pm$ 14.6	&-29 $\pm$ 16	&-12 $\pm$ 20	&{\bf12} $\pm$ 0.3 	&\nodata\\
G-21	&7	&{\bf140.7} $\pm$ 1.0	&{\bf140.1} $\pm$ 2.6	&267.6 $\pm$ 4.8	&{\bf-125} $\pm$ 13	&{\bf-165} $\pm$ 15	&{\bf271} $\pm$2.7	 &\nodata  \\
G-24	&13	&{\bf13.1} $\pm$ 0.1	&{\bf479.2} $\pm$ 9.4	&{\bf236.3} $\pm$ 16.5	&3 $\pm$ 11	&4 $\pm$ 14	& \nodata 	&{\bf72.8} $\pm$ 15.4 \\
\tableline								
Terrestrial	&-	&89	&499\tablenotemark{*}	&272\tablenotemark{*}	&\iso{29}{Si}/\iso{28}{Si} = .051	&\iso{30}{Si}/\iso{28}{Si} = .034	&0.05\tablenotemark{\dag\dag}	&0 \\
\enddata
\tablenotetext{\dag}{Delta-value notation,  \delval{i}{Si} = 1000$\times$\big((\iso{i}{Si}/\iso{28}{Si})$_{sample}$/(\iso{i}{Si}/\iso{28}{Si})$_{standard}$ - 1\big), deviations in permil from standard/terrestrial values.}
\tablenotetext{\dag\dag}{Inferred \iso{26}{Al}/\iso{27}{Al} ratio in Calcium-Aluminum-rich Inclusions, the oldest solar system solids.}
\tablenotetext{*}{The results of NASA's Genesis mission have shown that terrestrial materials are generally depleted in \iso{14}{N} and \iso{16}{O} relative to the solar wind. Solar wind isotopic ratios were found to be \iso{14}{N}/\iso{15}{N} = 441 \citep{marty2011} and \iso{16}{O}/\iso{18}{O} = 530 \citep{mckeegan2011}.}
\tablecomments{All values represent whole-grain measurements; anomalies are lower bounds due to surface-adhered insoluble organic material.  All errors are 1$\sigma$. Bold values indicate 2$\sigma$+ deviation from terrestrial.  Blank cells indicate no measurable excess in \iso{26}{Mg} or \iso{44}{Ca}.}
\end{deluxetable}
\clearpage

\end{document}